\documentclass[reprint,prb,aps, longbibliography]{revtex4-2}
\usepackage[english]{babel}
\usepackage{amsmath}
\usepackage{graphicx}
\usepackage{comment}
\usepackage{ulem}
\usepackage{xcolor}

\newcommand{\rvm}{\underset{=}{{\hspace{1}}\ensuremath{\mathbf{r}}}}

\def\be{\begin{equation}}
\def\ee{\end{equation}}
\def\ber{\begin{eqnarray}}
\def\eer{\end{eqnarray}}
\def\bern{\begin{eqnarray*}}
\def\eern{\end{eqnarray*}}

\def\rv{\ensuremath{\boldsymbol{r}}}

\def\Gv{\ensuremath{\boldsymbol{G}}}

\def\qv{\ensuremath{\boldsymbol{q}}}

\def\Pv{\ensuremath{\boldsymbol{P}}}
\def\Rv{\ensuremath{\boldsymbol{R}}}

\def\vv{\ensuremath{\boldsymbol{v}}}

\def\0v{\mathbf{0}}
\def\1v{\mathbf{1}}
\def\2v{\mathbf{2}}
\def\3v{\mathbf{3}}

\def\pa{\partial}

\DeclareMathAlphabet\mathbfcal{OMS}{cmsy}{b}{n}

\def\Im{ {\rm Im} \, }

\def\rvm{{\underset{\text{\raisebox{3 pt}{=}}}{r}}}

\begin{document}

\title{Many-body  interactions in the dielectric theory of stopping power of solids for classical and quantum projectiles}

\author{Vladimir~U.~Nazarov}
\affiliation{Fritz Haber Research Center for Molecular Dynamics and Institute of Chemistry, the  Hebrew University of Jerusalem, Jerusalem 9190401, Israel}
\email{vladimir.nazarov@mail.huji.ac.il}
%\affiliation{Research Center for Applied Sciences, Academia Sinica, Taipei 11529, Taiwan}

\author{Vyacheslav~M.~Silkin}
\affiliation{Donostia International Physics Center (DIPC), Paseo de Manuel Lardizabal 4, E-20018 San Sebasti\'an, Spain}
\affiliation{Departamento de Pol\'imeros y Materiales Avanzados: Física, Qu\'imica y Tecnolog\'ia, Facultad de Ciencias Qu\'imicas, Universidad del País Vasco (UPV-EHU), Apdo. 1072, E-20080 San Sebastián, Spain}
\affiliation{IKERBASQUE, Basque Foundation for Science, 48009 Bilbao, Spain}

\begin{abstract}
We take account of the many-body dynamic interactions by including the wave-vector and frequency-dependent exchange and correlations (xc) kernel $f_{xc}(\qv,\omega)$
in the framework of the dielectric theory of the stopping power of crystals  for moving charges. The cases of classical and quantum projectiles are considered.
We find that (I) the role of the xc effects in slowing of charges in crystalline  solids is more pronounced than it is within the jellium model and 
(II) For velocities below the stopping maximum, inclusion of xc leads to the improvement of the comparison of the theory with experiment over a range of solid targets. 
On the other hand, in the high-velocity regime, the role of the dynamic xc proves negligible,
which does not contradict experiment, and which we substantiate analytically within the jellium model of the target. 
The input to our theory is the microscopic dielectric matrix $\epsilon_{\Gv \Gv'}(\qv,\omega)$ within the random phase approximation (i.e., with  neglect of $f_{xc}$), 
the calculation of which matrix  is implemented in the existing solid-state codes.
\end{abstract}

\maketitle

\section{Introduction}

Interest in  slowing of ions travelling through matter dates back to the beginning of the last century \cite{Bragg-1905,Bohr-1913}. 
The modern theory of the stopping power (SP)  (energy loss by a projectile per unit pass length) had started from the works by Ritchie \cite{Ritchie-57,Ritchie-59}, 
wherein the target materials were modelled with the homogeneous electron gas (HEG), and the dielectric approach had been employed. With the advent of the density-functional theory (DFT) \cite{Hohenberg-64,Kohn-65} and the time-dependent DFT (TDDFT) \cite{Runge-84,Gross-85},
the limitations of the HEG model, as well as those of the first Born approximation, the latter underlying the dielectric approach, have been overstepped \cite{Echenique-86,Pruneda-07,Nazarov-05}.

The central quantities in  the real-time TDDFT and in its linear-response variant are the time-dependent exchange-correlation (xc) potential $v_{xc}(\rv,t)$ and the xc kernel $f_{xc}(\rv,\rv',t-t')$, respectively \cite{Runge-84,Gross-85}. These quantities are never known exactly, necessitating construction of approximations. 
Understandably, the more complicated is the system under study, the simpler and less accurate approximations are generally used for  $v_{xc}$ and $f_{xc}$. 
As a result, in the studies of SP of real solids, the adiabatic local density approximation (ALDA) to TDDFT 
is used almost exclusively, 
which choice suffers from the disadvantage of dropping out the effects due to the spatial and temporal non-locality of the electronic response of  target systems \cite{Gross-90,Giuliani&Vignale}. 

At the other end of the spectrum is an approach which employs the non-uniform jellium model and linear-response TDDFT, while striving for more accurate xc kernels \cite{Nazarov-05,Nazarov-07,Nazarov-08}. This method does preserve the non-locality, 
but it is applicable in the limit of zero velocity of the projectile only,  
the  quantity sought for being the friction coefficient $Q=\lim_{v\to 0} -\frac{1}{v} \frac{d E(v)}{d x}$. \cite{Roth-18}

In  the dielectric theory of SP, which will be further detailed below, 
the central assumption is the adoption of the perturbative treatment of the projectile-target interaction, to the first order in the latter (first Born approximation). 
In other respects, this theory is well fit to account for the more sophisticated features of the particle-solid interactions: 
(i) the static and dynamic xc can be naturally included in the dielectric formalism by means of the linear response TDDFT; (ii) the theory is designed for the case of finite velocities of the projectile; (iii) the theory does not rely on the mean-field approximation of the Ehrenfest dynamics \cite{Ehrenfest-27}, but it can be formulated totally quantum mechanically, not only with respect to the target system's electrons, but to the projectile particle as well 
(see Appendix \ref{App}); 
(iv) the theory is conceptually simple, requiring only the evaluation of the dielectric function (DF) of the target system.
On the other hand, while possessing an important advantage of the non-linearity,  other approaches to the problem of SP usually suffer from the lack of one or more of the features (i)-(iv).

The above observations make it essential to establish the bounds on the applicability of the dielectric approach for the SP problem, 
the ones which do not result from simplifications used in the construction of DF, but are rather inherent to the approach itself.
In this article, we undertake to implement  the dielectric theory of SP in its unabridged  form,
verifying  results against available experimental data.
At the level of the theory based on the accurate evaluation of the dielectric matrix of a periodic solid accompanied by the inclusion of the dynamic many-body (MB) interactions by means of the non-adiabatic linear-response TDDFT, we generally observe an improved comparison with experiment. Exceptions are also found and discussed.

This work is organized as follows. 
In Sec.~\ref{jel}, for the jellium solid model, we formulate, implement, and discuss the dielectric approach to the SP problem, which includes the quantum mechanical effects on the part of the projectile and the dynamical xc effects on the part of the target.
In Sec.~\ref{crys}, we perform the generalization of the results of Sec.~\ref{jel} to real solids.
Our conclusions are collected in Sec.~\ref{conc}.
More involved derivations and technical details are collected in Appendices \ref{App}-\ref{HV}.

\section{Jellium solid}
\label{jel}

Within the dielectric approach, SP for a  projectile  of the  charge $Z$ and infinite mass,  moving with the  velocity $v$ in a medium characterized by the wave-vector and frequency-dependent DF $\varepsilon(q,\omega)$ is given by \cite{Ritchie-57,Lindhard-64} 
\begin{equation} 
\frac{d E}{d x} = - \frac{2 Z^2}{\pi v^2 } \int\limits_0^{\infty} \frac{d q}{q} \int\limits_0^{v q} 
\omega  \, \Im \frac{1}{\epsilon\left( q, \omega \right)}  d \omega, 
\label{S1}
\end{equation}
wherein we use atomic units ($e^2=\hbar=m_e=1$). 
If the projectile has a finite mass $M$, then  the kinematic constraint   modifies Eq.~\eqref{S1} to \cite{Archubi-22,Archubi-23}
\begin{equation} 
\frac{d E}{d x} = - \frac{2 Z^2}{\pi v^2 } \int\limits_0^{2 M v} \frac{d q}{q} \int\limits_0^{v q-\frac{q^2}{2 M}} 
\omega \,  \Im \frac{1}{\epsilon\left( q, \omega \right)} d \omega.
\label{S2}
\end{equation}
From the outset,  it is conceptually important to differentiate between the classical and quantum mechanical features in Eqs.~\eqref{S1} and  \eqref{S2}.
As written in atomic units, Eq.~\eqref{S2}  disguises the distinction between the two kinds of effects. 
This distinction can be revealed by  re-introducing the world constants in Eq.~\eqref{S2}, while ensuring the proper dimensionalities
\begin{equation} 
\frac{d E}{d x} = - \frac{2 Z_1^2 e^2}{\pi v^2 } \int\limits_0^{2 M v/\hbar} \frac{d q}{q} \int\limits_0^{v q-\frac{\hbar q^2}{2 M}} 
\omega  \, \Im \frac{1}{\epsilon\left( q, \omega \right)} d \omega.
\label{S3}
\end{equation}
From Eq.~\eqref{S3}, it becomes clear  that the differences in the integration limits between Eqs.~\eqref{S1} and \eqref{S2} are due to quantum effects exclusively, since they disappear 
in the classical limit $\hbar\to  0$. 
In Appendix \ref{App} we give an independent derivation of the expression \eqref{QU} for SP of a periodic crystal, of which Eqs.~\eqref{S2} -- \eqref{S3} are a specific case.

For the jellium model of a solid,  Eqs.~\eqref{S1} and \eqref{S2} have been extensively used in the literature within the random phase approximation (RPA) to DF, which amounts to the neglect of xc effects \cite{Ritchie-57,Ritchie-59,Lindhard-64,Archubi-23}.
Our interest lying in  the role of the latter effects, 
we start from considering HEG of interacting electrons (electron liquid), taking into account the xc kernel $f_{xc}^h(q,\omega)$, which enters the expression for DF through the relations \cite{Gross-85}
\begin{align}
&\frac{1}{\epsilon(q,\omega)}=1+\frac{4\pi}{q^2} \chi^h(q,\omega), \label{KS1} \\
&\frac{1}{\chi^h(q,\omega)}=\frac{1}{\chi_s^h(q,\omega)} -\frac{4\pi}{q^2} - f_{xc}^h(q,\omega), \label{KS2}
\end{align}
where $\chi^h(q,\omega)$ and $\chi_s^h(q,\omega)$ are the interacting physical and the auxiliary independent-electron  (Kohn--Sham) density response functions, respectively,
the latter being known exactly by the Lindhard formula \cite{Lindhard-54}. 
The property of {\it viscosity} is  describable in terms of the frequency-dependent xc kernel $f_{xc}^h(q,\omega)$ \cite{Giuliani&Vignale},
which is especially important in the context of the transport phenomena \cite{Nazarov-24,Nazarov-26}.

\begin{figure}[h!]
\includegraphics[width=\columnwidth, clip=true, trim=58 3 2 7]{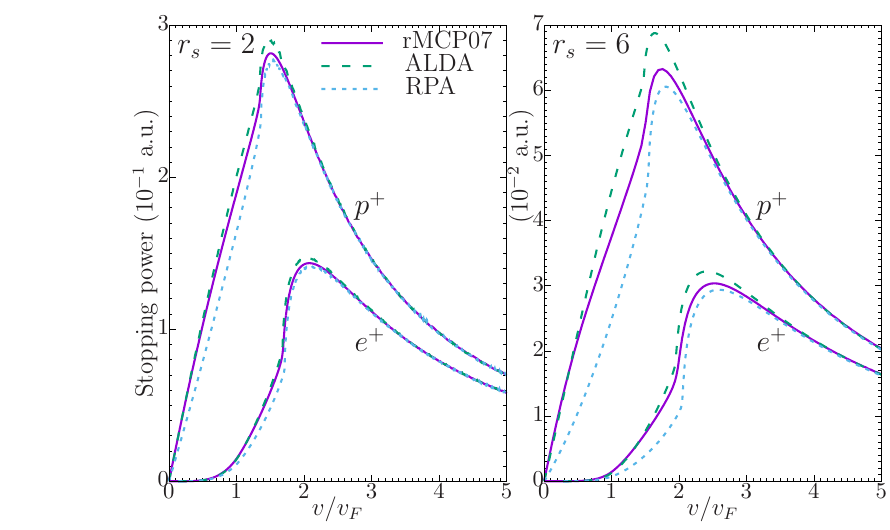}
\caption{\label{rs2_6}
Stopping power of jellium of the density parameter $r_s=2$ (left) and $6$ a.u. versus the velocity of the projectile  in units of the Fermi velocity $v_F$. 
Results for proton and positron, with the inclusion of the xc effects on the level of rMPC07 and ALDA (solid and dashed lines, respectively) and without xc (RPA, dotted lines) are shown.}
\end{figure}

\begin{figure}[h!]
\includegraphics[width=\columnwidth, clip=true, trim=48 3 2 7]{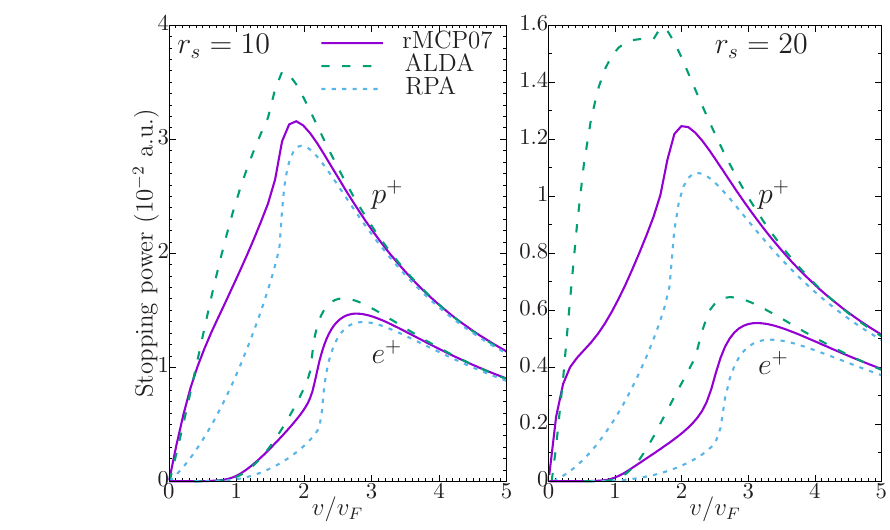}
\caption{\label{rs10_20}
Same as Fig.~\ref{rs2_6}, but for $r_s=10$ (left) and $20$ a.u.}
\end{figure}

With Eqs.~\eqref{S2}, \eqref{KS1}--\eqref{KS2}, we have conducted calculations of SP  of jellium of various densities for protons and positrons, within RPA ($f_{xc}^h=0$) and with xc effects included. For $f_{xc}^h$, we were using
the ALDA  and the recent {\it constraint-based revised modified} Constantin -- Pitarke (rMCP07) xc kernel \cite{Kaplan-22}, which is currently considered accurate at all densities of the fluid phase of HEG.
In Figs.~\ref{rs2_6} and \ref{rs10_20}, results for $r_s=2$, $6$ a.u. and $r_s=10$, $20$ a.u., respectively, where $r_s=[3/(4 \pi n)]^{1/3}$ is the density parameter of HEG of the density $n$, are shown. 
We observe that, at lower velocities, RPA systematically underestimates SP, while ALDA overestimates it, as compared with the calculation based on the rMCP07 parametrization.
This discrepancy grows with the decrease of the electronic  density (increase of $r_s$).
We note, and this applies to all the results below, that for proton SP is numerically indistinguishable from that for an infinite mass particle.

A notable feature in Figs.~\ref{rs2_6} and \ref{rs10_20} is the negligible contribution of $f_{xc}$, both rMCP07 and ALDA ones, in the higher-velocity regime. While this trend in the velocity-dependence of SP is far from obvious {\it a priori}, in Appendix \ref{HV} we prove that this holds for an arbitrary causal xc kernel. 

\begin{figure}[h!]
\includegraphics[width=\columnwidth, clip=true, trim=48 5 15 7]{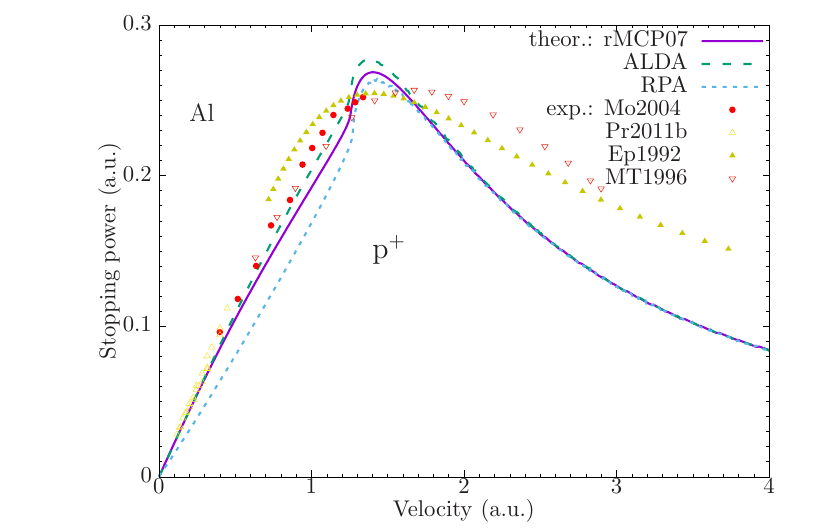}
\caption{\label{Al_te}
Stopping power of jellium of the density parameter $r_s=2.07$ a.u. for protons  calculated within the rMCP07 (solid line), ALDA (dashed line), and RPA (dotted line) approximations.
Symbols are  experimental data for the aluminium target: Mo2004 \cite{Moller-04}, Pr2001b \cite{Primetzhofer-11}, MT1996 \cite{Martinez-96}, and Ep1992 \cite{Eppacher-92}, as compiled in Ref.~\cite{Montanari-17-0,*Montanari-17}. The datasets' namings are those from the latter reference.
}
\end{figure}

\begin{figure}[h!]
\includegraphics[width=\columnwidth, clip=true, trim=48 5 19 8]{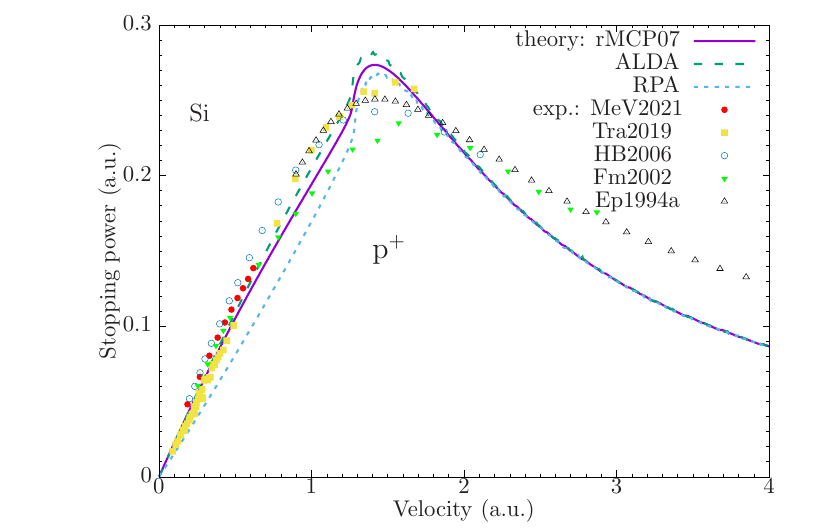}
\caption{\label{Si_te}
Same as Fig.~\ref{Al_te}, but for jellium of the density parameter $r_s=2.05$ a.u. Symbols are  experimental data for the silicon target: MeV2021 \cite{Mery-21}, Tra2019 \cite{Tran-19}, HB2006 \cite{Hobler-06}, Fm2002 \cite{Fama-02}, and Ep1994a \cite{Eppacher-95}, as compiled in Ref.~\cite{Montanari-17-0,*Montanari-17}.
}
\end{figure}

In Figs.~\ref{Al_te}--\ref{Si_te}, we compare the jellium model calculation results with experimental SP data on  aluminium and silicon targets for protons.
On the left slope of the peaks (lower velocities), we observe an improved  agreement with experiment due to the account for the xc effects. 
We note that ALDA performs slightly better than rMCP07, which, however, is an artefact of neglecting the crystallinity effects, as will be seen below.

To the right from the maximum (higher velocities) the theory largely underestimates the experimental SP.
It is known that, at the growing velocities, SP becomes strongly affected by the excitation of the core electrons \cite{Shukri-16,Archubi-23}, which effect the jellium model cannot, obviously, capture.

The role of  xc in the SP of HEG for classical projectiles has been earlier addressed in the extensive literature \cite{Tanaka-85,Wang-97,Barriga-Carrasco-08,Barriga-Carrasco-09}. Those studies, accordingly, used earlier available parametrizations for the local field correction function (or, equivalently, xc kernel) \cite{Ichimaru-81,Utsumi-82,Ichimaru-85-1,Dabrowski-86}, or have used Mermin's dielectric function, as in Ref.~\cite{Barriga-Carrasco-08}. 
These have been surpassed in the more recent developments \cite{Ruzsinszky-20,Kaplan-22}. 
We also note a promising first-principles (not constraint-based) exchange-only kernel, derived from a time-dependent variational principle  \cite{Nazarov-13-2,Nazarov-24-2}.
Here we are using the state-of-the-art exchange and correlation parametrization rMCP07 \cite{Kaplan-22}.

\section{Crystals}
\label{crys}

We refine the theory by including the crystallinity effects in real solids. The  generalization of Eq.~\eqref{S2} for a spatially periodic system and random SP is
(see Appendix \ref{App})
\begin{equation} 
\begin{split}
&\frac{d E}{d x}  = - \frac{ Z^2}{ \pi^2 v } \int\limits_{\qv\in \Omega_B} d\qv \sum\limits_{\Gv}  
  \frac{\vv\cdot (\qv+\Gv) -\frac{(\qv+\Gv)^2}{2 M}}{|\qv+\Gv|^2} \times \\
& \Theta \! \! \left[\vv \! \cdot \! (\qv \! + \! \Gv) \! - \! \frac{(\qv \! + \! \Gv)^2}{2 M}\right]
       \!  \Im  \epsilon^{-1}_{\Gv \Gv} \! \left[\qv,\vv \! \cdot \! (\qv \! + \! \Gv) \! - \! \frac{(\qv \! + \! \Gv)^2}{2 M}\right] \! \! ,
\end{split} 
\label{QU}
\end{equation}
where $\Theta$ is the Heaviside step function.
In Eq.~\eqref{QU} the summation is over the infinite set of the reciprocal vectors $\Gv$ of the lattice, 
the integration in $\qv$ is over the first Brillouin zone,
and $\epsilon^{-1}_{\Gv \Gv'}(\qv,\omega)$ is the inverse  microscopic dielectric matrix of the system, 
of which the diagonal elements only are needed.
Equation~\eqref{QU} is the quantum mechanical generalization of the corresponding classical SP formula for periodic solids
\cite{Campillo-98,Shukri-16}.

Within the linear-response TDDFT of crystals, the matrix version of the relations \eqref{KS1} -- \eqref{KS2} reads
\begin{align}
&\epsilon^{-1}_{\Gv \Gv'}(\qv,\omega)=1+\frac{4\pi}{|\qv+\Gv| |\qv+\Gv'|} \chi_{\Gv \Gv'}(\qv,\omega), \label{KS1c} \\
&\chi^{-1}_{\Gv \Gv'}(\qv,\omega) \! = \! \chi^{-1}_{s,\Gv \Gv'}(\qv,\omega) \!-\!\frac{4\pi}{|\qv  + \Gv|^2} \delta_{\Gv \Gv'} \! - \! f_{xc,\Gv \Gv'}(\qv,\omega) \label{KS2c}.
\end{align}

We were calculating $\chi_{s,\Gv \Gv'}(\qv,\omega)$ with the all-electron full-potential linearised augmented-plane wave code ELK \cite{elk}.
At larger projectiles' velocities,  the excitation of the core states of the target plays the major role \cite{Shukri-16,Archubi-23},
which, in the calculations, necessitates the promotion of the core to valence electrons and, as a consequence, the use of large dielectric matrices.
Further particulars of the  calculations are collected in Appendix \ref{DC}.

The spatial and temporal non-locality of $f_{xc}$ was accounted for within the approximation of HEG 
with the average, over the unit cell, density of valence electrons, i.e.,
\begin{equation}
f_{xc,\Gv \Gv'}(\qv,\omega) \approx f^h_{xc}(\qv+\Gv,\omega) \delta_{\Gv \Gv'},
\label{fxc_HEG}
\end{equation}
where for $f^h_{xc}$ the  parametrization rMCP07 was used \cite{Kaplan-22}.

\begin{figure}[h!]
\includegraphics[width=\columnwidth, clip=true, trim=48 5 19 8]{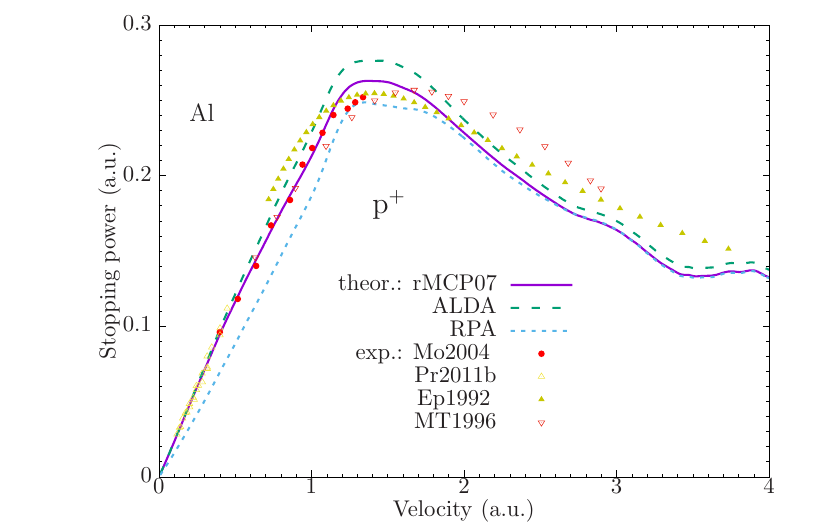}
\caption{\label{Al_te_elk}
Stopping power of aluminium crystal for protons with (rMCP07 solid line, ALDA dashed line) and without (RPA, dotted line) inclusion  of the xc effects. 
Symbols are  experimental data: Mo2004 \cite{Moller-04}, Pr2001b \cite{Primetzhofer-11}, MT1996 \cite{Martinez-96}, and Ep1992 \cite{Eppacher-92}, as compiled in Ref.~\cite{Montanari-17-0,*Montanari-17}. 
}
\end{figure}

In Fig.~\ref{Al_te_elk}, SP of crystalline aluminium, calculated with the use of Eqs.~\eqref{QU}--\eqref{fxc_HEG}, 
is plotted against the velocity of protons and compared with experimental data.
To the left of the peak, inclusion of the crystallinity together with xc effects makes a considerable difference, moving theoretical results closer to the measured data.
To the right from the peak, xc effects play a negligible role, while the real-solid calculation catches  partially excitations  from the deeper electron levels. 
The waviness of the spectra is due to the comparatively sparse $q$-points grid used for integration,
and it was confirmed to be smoothed out by making the grid denser, however, at the cost of much more expensive computations.

To take account of the excitations from the deeper electronic levels, a large number of $\Gv$-vectors must be used in Eq.~\eqref{QU},
owing to the localized character of those levels, which requires  many terms in their Fourier transform. 
In  Appendix \ref{DC} we list the parameters used in the calculations.
The further increase of the computational complexity was beyond our technical means, which contributes to the remaining discrepancies with experiment at the higher projectiles' velocities.

\begin{figure}[h!]
\includegraphics[width=\columnwidth, clip=true, trim=48 5 19 8]{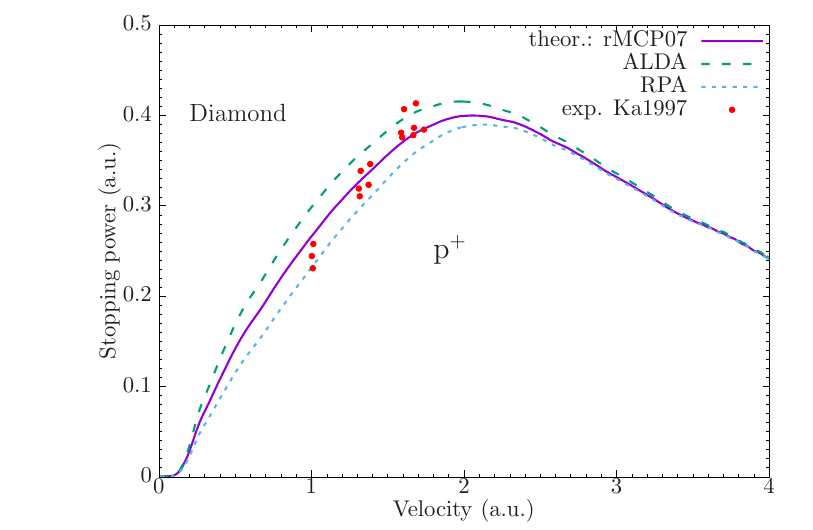}
\caption{\label{C_te_elk}
Stopping power of diamond crystal for protons with (solid line) and without (RPA, dashed line) inclusion  of the xc effects. Symbols are  experimental data  \cite{Kaferbock-97}, as compiled in Ref.~\cite{Montanari-17-0,*Montanari-17}.
}
\end{figure}

\begin{figure}[h!]
\includegraphics[width=\columnwidth, clip=true, trim=48 5 19 8]{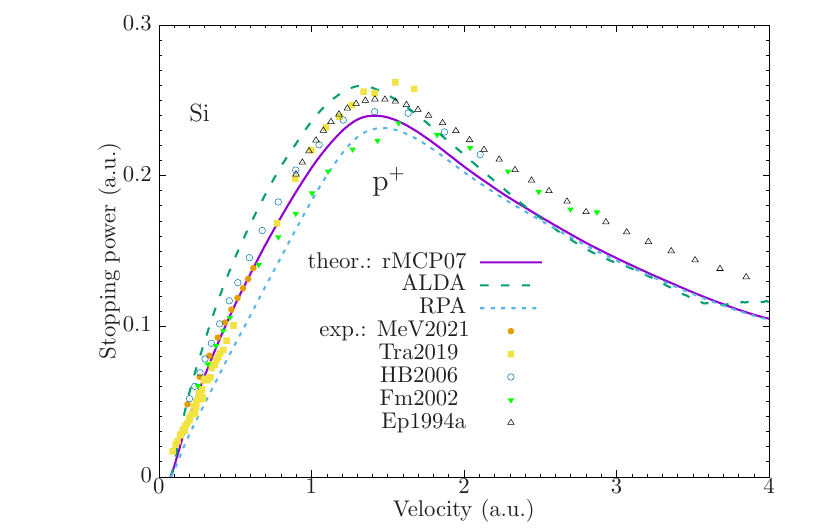}
\caption{\label{Si_te_elk}
Stopping power of silicon crystal for proton projectiles with (rMCP07 solid line, ALDA dashed line) and without (RPA, dotted line) inclusion  of the xc effects. 
Symbols are  experimental data: MeV2021 \cite{Mery-21}, Tra2019 \cite{Tran-19}, HB2006 \cite{Hobler-06}, Fm2002 \cite{Fama-02}, and Ep1994a \cite{Eppacher-95}, as compiled in Ref.~\cite{Montanari-17-0,*Montanari-17}.
}
\end{figure}

\begin{figure}[h!]
\includegraphics[width=\columnwidth, clip=true, trim=48 5 19 8]{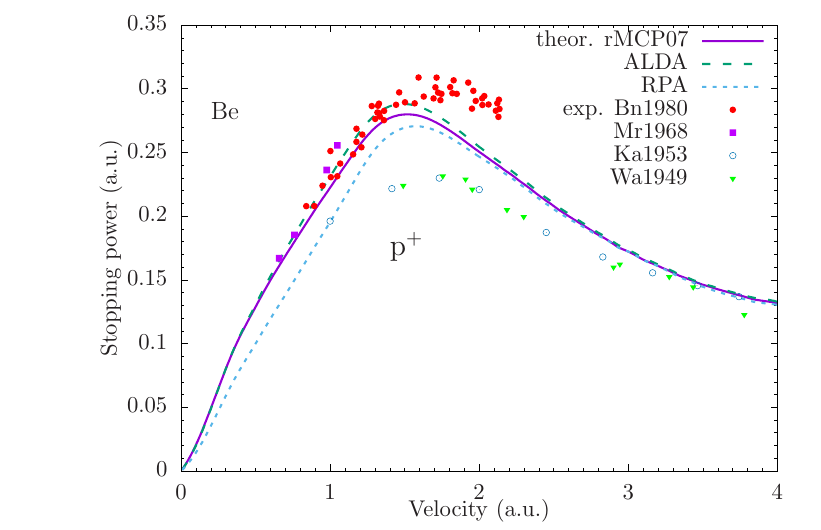}
\caption{\label{Be_te_elk}
Stopping power of beryllium crystal for protons with (rMCP07 solid line, ALDA dashed line) and without (RPA, dotted line)  inclusion  of the xc effects. Symbols are  experimental data: Bn1980 \cite{Brunner-80}, Mr1968 \cite{Morita-68}, Ka1953 \cite{Kahn-53}, and Wa1949 \cite{Warshaw-49}, as compiled in Ref.~\cite{Montanari-17-0,*Montanari-17}.
}
\end{figure}

Results for diamond, Figs.~\eqref{C_te_elk},   silicon, Fig.~\eqref{Si_te_elk}, and beryllium, Fig.~\eqref{Be_te_elk}, are qualitatively similar to those for aluminium,
clearly confirming the importance of the combined crystallinity and xc effects at and to the left from the spectra maxima.
The older experimental data for Be are, however, in dissonance with our theoretical, as well as with the newer experimental, results.

\begin{figure}[h!]
\includegraphics[width=\columnwidth, clip=true, trim=48 5 19 8]{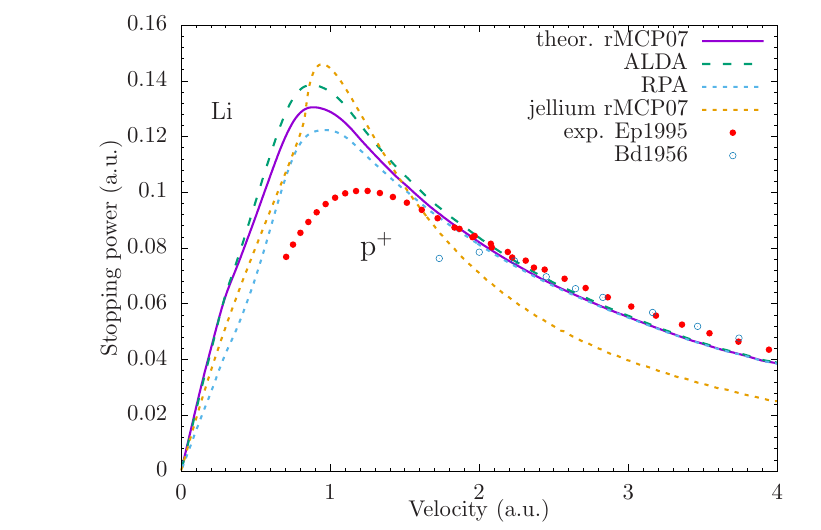}
\caption{\label{Li_te_elk}
Stopping power of lithium crystal for protons with (rMCP07 solid line, ALDA dashed line) and without (RPA, dotted line)  inclusion  of the xc effects. Symbols are  experimental data: Ep1995 \cite{Eppacher-95-2}, Bd1956 \cite{Bader-56}, as compiled in Ref.~\cite{Montanari-17-0,*Montanari-17}.
The jellium model results are also shown in order to contrast a difference with the real-solid calculation for this low-conduction-electron density metal.
}
\end{figure}

\begin{figure}[h!]
\includegraphics[width=\columnwidth, clip=true, trim=48 5 19 8]{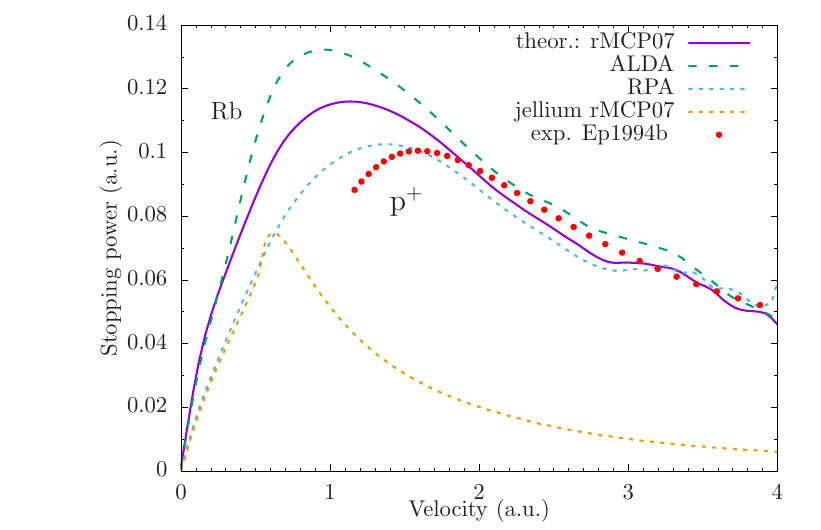}
\caption{\label{Rb_te_elk}
Stopping power of rubidium crystal for protons with (rMCP07 solid line, ALDA dashed line) and without (RPA, dotted line)  inclusion  of the xc effects. Symbols are  experimental data: Ep1994b \cite{Eppacher-94}, as compiled in Ref.~\cite{Montanari-17-0,*Montanari-17}.
The calculated curves have been smoothed to eliminate a strong noise.
The jellium model results are also shown in order to contrast a drastic difference with the real-solid calculation for this low-conduction-electron density metal.
}
\end{figure}

Alkali metals lithium and rubidium, Figs~~\ref{Li_te_elk} and \ref{Rb_te_elk}, respectively, present notable exceptions to the agreement between the theory and experiment at and to the left from the  main maxima.
The dielectric theory of SP, obviously, breaks down  for those metals in this velocity range.
This can be understood as follows: In the low velocity regime and at the lower densities of HEG,
a bound state of the pseudo-atom is formed \cite{Echenique-86}, which, for the proton projectile, happens at $r_s\gtrsim 2$ a.u.
The first Born approximation for the inelastic scattering cannot, obviously, capture this process, resulting in the failure of the dielectric approach
for Li ($r_s=3.25$ a.u.) and Rb ($r_s=5.32$ a.u). On the other hand, although for proton in the Al ($r_s=2.07$ a.u.) and Si ($r_s=2.05$ a.u.) hosts the bound state develops as well, it is marginally shallow, practically not affecting the SP.

To the right from the peaks the comparison is fairly good.
We, further, note that Rb turns out an example of a simple metal for which the jellium model is irrelevant at all to the SP calculation, as  seen from Fig.~\ref{Rb_te_elk}.
Our choice of alkali targets was determined by the availability of experimental data \cite{Montanari-17}.

\begin{figure}[h!]
\includegraphics[width=\columnwidth, clip=true, trim=47 5 19 7]{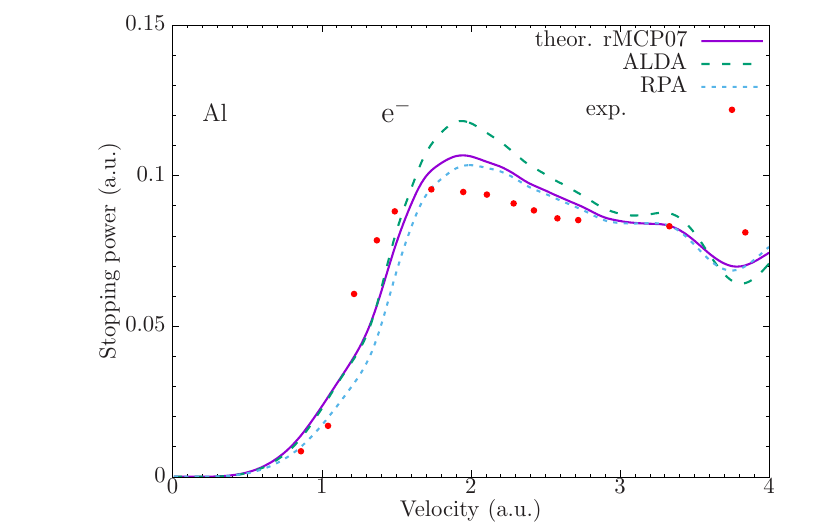}
\caption{\label{Al_electron}
Stopping power of aluminium crystal for electron with (rMCP07 solid line, ALDA dashed line) and without (RPA, dotted line)  inclusion  of the xc effects. 
Symbols are  experimental data from Ref.~\cite{Joy-96}.
}
\end{figure}

\begin{figure}[h!]
\includegraphics[width=\columnwidth, clip=true, trim=47 5 19 7]{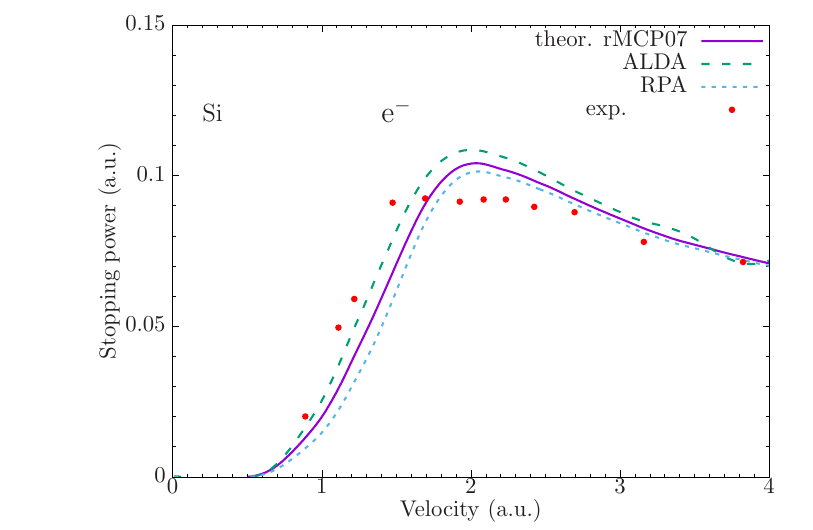}
\caption{\label{Si_electron}
Stopping power of silicon crystal for electron with (rMCP07 solid line, ALDA dashed line) and without (RPA, dotted line)  inclusion  of the xc effects. 
Symbols are  experimental data from Ref.~\cite{Joy-96}.
}
\end{figure}

In Figs.~\ref{Al_electron} and \ref{Si_electron}, the dependence of the stopping power of Al and Si on the velocity is plotted for electron projectiles.
The factor
\begin{equation}
g_x=1+\frac{|\qv+\Gv|^4}{v^4}-\frac{|\qv+\Gv|^2}{v^2},
\end{equation} 
which describes the exchange between the impinging electron and those of the target \cite{Ochkur-64}, was included under the sum in Eq.~\eqref{QU}.
The overall agreement between the theory and experiment is fairly good, although xc in the response of the target does not play a major role in the case of the electron projectile.
We note that the rapid disappearance of the SP for electron at lower velocities in Figs.~\ref{Al_electron} and \ref{Si_electron} is in accord with the results for the friction coefficient in the zero-velocity limit within the nonlinear theory \cite{Nazarov-26-3}.

\section{Conclusions}
\label{conc}

We have enhanced the dielectric theory of the stopping power of solids for moving charges by simultaneously accounting for the crystallinity and the  many-body interaction effects. The many-body effects have been included by means of the linear-response TDDFT via the dynamic exchange and correlation kernel as an ingredient to the dielectric response of a solid.

We have found the role of the many-body effects, in combination with the crystallinity ones, quite important. 
Moreover, the account for those effects brings the theory into an improved comparison with experiment for the majority of considered targets. 
At the same time, we have observed a  discrepancy between the theory and experiment for alkali metals (Li and Rb) at lower velocities of projectiles, which signifies the breakdown of the dielectric theory of the stopping power for those targets in the low velocity regime,
prompting for the resolution to be sought beyond the linear theory.

\acknowledgements
V.U.N.  acknowledges the hospitality of the Donostia International Physics Center, San Sebasti\'an, Spain.
V.M.S. acknowledges financial support by Grant PID2022-139230NB-I00 funded by MCIN/AEI/10.13039/501100011033.

{\it Data availability} -- The data that support the findings of
this article are openly available 
\footnote{V.~U.~Nazarov and V.~M.~Silkin, Many-body  interactions in the dielectric theory of stopping power of solids for classical and quantum projectiles, 
DOI: 10.6084/m9.figshare.32089053}.

%\bibliography{ref}


\appendix

\section{Quantum mechanical derivation of the formula for the stopping power within the dielectric formalism}

\label{App}

We consider a target system of $N_e$ electrons and $N_n$ nuclei. A projectile charge $Z$ is travelling through this system. 
We consider electrons of the target and the projectile as quantum particles, while nuclei of the target are point charges pinned at their equilibrium positions.
The time-dependent Schr\"{o}dinger equation for the wave-function $\Pi(\rvm,\Rv,t)$ of the whole system, where $\rvm$ stands for the set of the electronic coordinates $\rv_i$, reads
\begin{equation}
i \frac{\pa \Pi(\rvm,\Rv,t)}{\pa t} = \hat{H} \Pi(\rvm,\Rv,t),
\end{equation}
where
\begin{equation} \hat{H} = \hat{H}_0 + \frac{\hat{\Pv}^2}{2 M} + \hat{H}_1 , \end{equation}
$\hat{H}_0$ is the Hamiltonian of the target, $\hat{\Pv}=-i \nabla_{\Rv}$ is the momentum operator of the projectile, $M$ being its mass,
\begin{equation} 
\hat{H}_1 = \sum\limits_{i = 1}^{N_e} - \frac{Z}{|\rv_i - \Rv|} +\sum\limits_{j = 1}^{N_n}  \frac{Z_j Z}{|\Rv_j - \Rv|} , 
\label{H1}
\end{equation}
where $\Rv_j$ and $Z_j$ are the coordinates and charges of nuclei, respectively, and  $\Rv$ is the coordinate of the projectile.

Let, at $t\to - \infty$, the target and the projectile wave-functions be well separated, and the target to be in its ground-state $\Psi_0(\rvm)$.
Then we can write to the first Born approximation in $\hat{H}_1$
\onecolumngrid
\begin{equation} 
\begin{split}
&\tilde{\Pi} (\rvm,\Rv,t\to +\infty) = \Psi_0(\rvm)  \tilde{\Phi}(\Rv,t\to -\infty) + \frac{1}{i} 
\int_{-
   \infty}^{\infty} \! \! \! d t e^{i \hat{H}_0 t} e^{i t \hat{\Pv}^2 / 2 M} 
   \hat{H}_1 e^{- i \hat{H}_0 t} e^{- i t \hat{\Pv}^2 / 2 M} \Psi_0(\rvm) 
   \tilde{\Phi}(\Rv,t\to \! -\infty),
\end{split}
\end{equation}
where the overhead tilde denotes the corresponding wave-function in the interaction picture. Furthermore,      
\begin{equation} 
\begin{split}
&\left\langle \Psi_{n \neq 0}(\rvm)  \left| \tilde{\Pi} (\rvm,\Rv,t\to +\infty) \right.\right\rangle_\rvm \! \! =
   \frac{1}{i}
\int_{- \infty}^{\infty} \! \! \! \!  d t \, e^{i (E_n - E_0) t} e^{i t
   \hat{\Pv}^2 / 2 M} 
 \langle \Psi_n(\rvm) | \hat{H}_1 | \Psi_0(\rvm) \rangle_\rvm
   e^{- i t \hat{\Pv}^2 / 2 M}  \tilde{\Phi}(\Rv,t\to -\infty) ,
   \label{E5}
\end{split}
\end{equation}
where $\Psi_n$ and $E_n$ are the complete set of the eigenfunctions of the target and its corresponding eigenenergies, respectively, 
and $\langle \dots \rangle_\rvm$ denotes integration over electronic coordinates.
Substituting Eq.~\eqref{H1} into Eq.~\eqref{E5}, introducing the electron density operator
\begin{equation}
\hat{n}(\rv)=\sum\limits_{i=1}^{N_e} \delta(\rv_i-\rv),
\end{equation}
and noting that the second term in Eq.~\eqref{H1} does not contribute to the matrix element, we can rewrite Eq.~\eqref{E5} as
%\begin{widetext}
\begin{equation} 
\left\langle \Psi_{n \neq 0}(\rvm)  \left| \tilde{\Pi} (\rvm,\Rv,t\to +\infty)\right.  \right\rangle_{\! \rvm} \! = \!
   \frac{Z}{i}  \int \! \! \left\langle \Psi_n(\rvm)  \left| \hat{n} (\rv) \right|
   \Psi_0(\rvm)  \right\rangle_{\! \rvm} \! \! \! e^{i (E_n - E_0) t} e^{it \hat{\Pv}^2 / 2 M} 
   \frac{1}{|\rv-\Rv|} e^{- it \hat{\Pv}^2 / 2 M} 
   \tilde{\Phi}(\Rv,t\to-\infty) d\rv dt,
    \end{equation}
%\end{widetext}
or, after the Fourier transform
\begin{equation} \tilde{\Phi} (\Pv,t) = (2 \pi)^{- 3 / 2}  \int \tilde{\Phi}
   (\Rv,t) e^{-i \Pv \cdot \Rv} d\Rv,
\end{equation}  
%\begin{widetext}
\begin{equation} 
\begin{split}
\left\langle \Psi_{n \neq 0}(\rvm)  \left| \tilde{\Pi} (\rvm,\Rv,t\to +\infty) \right. \right\rangle_{\! \rvm} &=
   \frac{Z}{ i \pi (2 \pi)^{5 / 2}}  \int \left\langle \Psi_n(\rvm)  \left|
   \hat{n} (\rv) \right| \Psi_0(\rvm)  \right\rangle_{\! \rvm} e^{i (E_n - E_0) t}
   e^{it (\Pv-\qv)^2 / 2 M} \\
&\times   
   e^{i\qv \cdot \rv} e^{- it\Pv^2 / 2 M}  \frac{d\qv}{q^2} 
   \tilde{\Phi} (\Pv,t\to-\infty) e^{i (\Pv-\qv) \cdot
   \Rv} d\Pv d\rv dt . 
   \end{split}
   \label{Bd}
    \end{equation}
%\end{widetext}
Integrating in Eq.~\eqref{Bd} over time explicitly, we can write 
%\begin{widetext} 
\begin{equation} 
\begin{split}
\left\langle \Psi_{n \neq 0}(\rvm)  \left| \tilde{\Pi} (\rvm,\Rv,t\to+\infty) \right. \right\rangle_{\!\rvm} &=
   \frac{Z}{i \pi (2 \pi)^{3 / 2}}  \int \left\langle \Psi_n(\rvm)  \left| \hat{n}
   (\rv) \right| \Psi_0(\rvm)  \right\rangle_{\!\rvm} \delta \left[ E_n - E_0 +
   \frac{(\Pv-\qv)^2}{2 M} - \frac{\Pv^2}{2 M}
   \right] \\
   &\times e^{i\qv \cdot \rv}  \frac{d\qv}{q^2} 
   \tilde{\Phi} (\Pv,t\to-\infty) e^{i (\Pv-\qv) \cdot
   \Rv} d\Pv d\rv ,
\end{split}   
\label{S10}
\end{equation}
and
\begin{equation} 
\begin{split}
&\int \left| \left\langle \Psi_{n \neq 0}(\rvm)  \left| \tilde{\Pi} (\rvm,\Rv,t\to+\infty) \right. \right\rangle_{\!\rvm} \right|^2 \! d \Rv =
   \frac{Z^2}{8 \pi^5 }  \int  \delta \left[ E_n - E_0 +
   \frac{(\Pv-\qv)^2}{2 M} - \frac{\Pv^2}{2 M}
   \right] \\
   &\times e^{i\qv \cdot \rv}  
   e^{i (\Pv-\qv) \cdot
   \Rv} 
\left\langle \Psi_n(\rvm)  \left| \hat{n}
   (\rv) \right| \Psi_0(\rvm)  \right\rangle_{\!\rvm} \left\langle \Psi_n(\rvm)  \left| \hat{n}
   (\rv') \right| \Psi_0(\rvm)  \right\rangle_{\!\rvm}^{\!*} \delta \left[ E_n - E_0 +
   \frac{(\Pv'-\qv')^2}{2 M} - \frac{{\Pv'}^2}{2 M}
   \right] \\
   &\times e^{-i\qv' \cdot \rv'}  e^{-i (\Pv'-\qv')\cdot\Rv} 
  \tilde{\Phi} (\Pv,t\to-\infty)  \tilde{\Phi}^* (\Pv',t\to-\infty)   \frac{d\qv d\qv'}{q^2 {q'}^2} d\Pv d\rv d\Pv' d\rv' d\Rv,
\end{split}   
\label{S101}
\end{equation}
or, integrating further over $\Rv$ and $\Pv'$,
\begin{equation} 
\begin{split}
&\int \left| \left\langle \Psi_{n \neq 0}(\rvm)  \left| \tilde{\Pi} (\rvm,\Rv,t\to+\infty) \right. \right\rangle_{\!\rvm} \right|^2 \! d \Rv =
   \frac{Z^2}{\pi^2 }  \int  \delta \left[ E_n - E_0 +
   \frac{(\Pv-\qv)^2}{2 M} - \frac{\Pv^2}{2 M}
   \right] \\ 
&\times \left\langle \Psi_n(\rvm)  \left| \hat{n}
   (\rv) \right| \Psi_0(\rvm)  \right\rangle_{\!\rvm} \left\langle \Psi_n(\rvm)  \left| \hat{n}
   (\rv') \right| \Psi_0(\rvm)  \right\rangle_{\!\rvm}^{\!*} \delta \left[  \frac{\Pv^2}{2 M}- \frac{(\Pv-\qv+\qv')^2}{2 M}
   \right] \\
 &  \times e^{i \qv\cdot \rv-i\qv' \cdot \rv'} 
  \tilde{\Phi} (\Pv,t\to-\infty)  \tilde{\Phi}^* (\Pv-\qv+\qv',t\to-\infty)   \frac{d\qv d\qv'}{q^2 {q'}^2} d\Pv d\rv  d\rv' .
\end{split}   
\label{S12}
\end{equation}
%\end{widetext}

For the energy gain by the target after the interaction we can write
\begin{equation} 
\Delta E = \sum_{n\ne 0} (E_n - E_0)  \int \left |\langle \Psi_n(\rvm) | \tilde{\Pi} (\rvm,\Rv,t\to+\infty) \rangle_\rvm  \right|^2 d\Rv,  
\label{dE}  
\end{equation}
which, with the use of Eq.~\eqref{S12}, yields
%\begin{widetext}
\begin{equation} 
\begin{split}
\Delta E &= \frac{Z^2}{\pi^2 }  \int \sum_{n\ne 0} \left[\frac{\Pv^2}{2 M}-\frac{(\Pv-\qv)^2}{2 M} \right]  \delta \left[  \frac{\Pv^2}{2 M}- \frac{(\Pv-\qv+\qv')^2}{2 M}
   \right]
 \delta \left[ E_n - E_0 +
   \frac{(\Pv-\qv)^2}{2 M} - \frac{\Pv^2}{2 M}
   \right] \\ 
&\times \left\langle \Psi_n(\rvm)  \left| \hat{n}
   (\rv) \right| \Psi_0(\rvm)  \right\rangle_{\!\rvm} 
   \left\langle  \Psi_0(\rvm)  \left| \hat{n}   (\rv') \right|\Psi_n(\rvm)  \right\rangle_{\!\rvm} 
  e^{i \qv\cdot \rv-i\qv' \cdot \rv'} 
  \tilde{\Phi} (\Pv,t\to-\infty)  \tilde{\Phi}^* (\Pv-\qv+\qv',t\to-\infty)   \frac{d\qv d\qv'}{q^2 {q'}^2} d\Pv d\rv  d\rv'  .
\end{split}
\label{dE2}  
\end{equation}

On the other hand, for the density response function  $\chi (\rv, \rv', \omega)$ of the same target one can write \cite{Giuliani&Vignale}
\begin{equation}
  \chi (\rv, \rv', \omega) = \sum_{n \ne 0} \left[
  \frac{\langle \Psi_n(\rvm) | \hat{n} (\rv) | \Psi_0(\rvm) \rangle_\rvm \langle
  \Psi_0(\rvm) | \hat{n} (\rv') | \Psi_n(\rvm) \rangle_\rvm }{E_0 - E_n + \omega + i
  \eta} + \frac{\langle \Psi_n(\rvm) | \hat{n} (\rv') | \Psi_0(\rvm) \rangle_\rvm
  \langle \Psi_0(\rvm) | \hat{n} (\rv) | \Psi_n(\rvm) \rangle_\rvm}{E_0 - E_n -
  \omega - i \eta} \right], \label{chi}
\end{equation}
%\end{widetext}
and, therefore,
\begin{equation} 
\Theta (\omega) \Im \chi (\rv, \rv', \omega) = -
   \pi \sum_{n \ne 0} \langle \Psi_n(\rvm) | \hat{n} (\rv) | \Psi_0(\rvm)
   \rangle_\rvm  \langle \Psi_0(\rvm) | \hat{n} (\rv') | \Psi_n(\rvm) \rangle_\rvm \delta
   (E_n - E_0 - \omega), 
\label{Imchi}
\end{equation}
where $\Theta (\omega)$ is the Heaviside step-function. Combining Eqs.~\eqref{dE2} and \eqref{Imchi}, we have
\begin{equation} 
\begin{split}
\Delta E &= -\frac{Z^2}{\pi^3 }  \int  \left[\frac{\Pv^2}{2 M}-\frac{(\Pv-\qv)^2}{2 M} \right] \Theta\left[\frac{\Pv^2}{2 M}-\frac{(\Pv-\qv)^2}{2 M} \right] 
 \delta \left[  \frac{\Pv^2}{2 M}- \frac{(\Pv-\qv+\qv')^2}{2 M}
   \right]
 \Im  \chi\left[\rv,\rv',\frac{\Pv^2}{2 M}-\frac{(\Pv-\qv)^2}{2 M}\right]
    \\ 
&\times 
  e^{i \qv\cdot \rv-i\qv' \cdot \rv'} 
  \tilde{\Phi} (\Pv,t\to-\infty)  \tilde{\Phi}^* (\Pv-\qv+\qv',t\to-\infty)   \frac{d\qv d\qv'}{q^2 {q'}^2} d\Pv d\rv  d\rv'  .
\end{split}
\label{dE3}  
\end{equation}
   
In the case of a crystal the density response function $\chi(\rv,\rv',\omega)$ can be written in terms of the matrix $\chi_{\Gv \Gv'}(\qv,\omega)$ indexed with the reciprocal vectors of the crystal lattice
\begin{equation}
\chi(\rv,\rv',\omega)= \frac{1}{(2\pi)^3} \int\limits_{\qv\in \Omega_B} d\qv \sum\limits_{\Gv,\Gv'} \chi_{\Gv \Gv'}(\qv,\omega)
e^{i \qv\cdot (\rv-\rv')} e^{i \Gv\cdot\rv-i \Gv'\cdot \rv'},
\end{equation}   
where $\Omega_B$ is the first Brillouin zone. Therefore,
\begin{equation} 
\begin{split}
\Delta E &= - \frac{Z^2}{\pi^3 (2\pi)^3} \int\limits_{\qv''\in \Omega_B} d\qv'' \sum\limits_{\Gv,\Gv'}  \int  \left[\frac{\Pv^2}{2 M}-\frac{(\Pv-\qv)^2}{2 M} \right] \Theta\left[\frac{\Pv^2}{2 M}-\frac{(\Pv-\qv)^2}{2 M} \right] 
 \delta \left[  \frac{\Pv^2}{2 M}- \frac{(\Pv-\qv+\qv')^2}{2 M}
   \right]  e^{i \qv\cdot \rv-i\qv' \cdot \rv'} 
    \\ 
&\times 
  \tilde{\Phi} (\Pv,t\to-\infty)  \tilde{\Phi}^* (\Pv-\qv+\qv',t\to-\infty)   \frac{d\qv d\qv'}{q^2 {q'}^2} d\Pv d\rv  d\rv' 
\Im  \chi_{\Gv \Gv'}\left[\qv'',\frac{\Pv^2}{2 M}-\frac{(\Pv-\qv)^2}{2 M}\right]
e^{i \qv''\cdot (\rv-\rv')} e^{i \Gv\cdot\rv-i \Gv'\cdot \rv'}. 
\end{split} 
\label{int0}
\end{equation}
In Eq.~\eqref{int0} we first integrate in $\rv'$ over the whole space
\begin{equation} 
\begin{split}
\Delta E &= - \frac{Z^2}{\pi^3 } \int\limits_{\qv''\in \Omega_B} d\qv'' \sum\limits_{\Gv,\Gv'}  \int  \left[\frac{\Pv^2}{2 M}-\frac{(\Pv-\qv)^2}{2 M} \right] \Theta\left[\frac{\Pv^2}{2 M}-\frac{(\Pv-\qv)^2}{2 M} \right] 
 \delta \left[  \frac{\Pv^2}{2 M}- \frac{(\Pv-\qv-\qv''-\Gv')^2}{2 M}
   \right]  e^{i (\qv+\qv''+\Gv)\cdot \rv} 
    \\ 
&\times 
  \tilde{\Phi} (\Pv,t\to-\infty)  \tilde{\Phi}^* (\Pv-\qv-\qv''-\Gv',t\to-\infty)   \frac{d\qv d\Pv }{q^2 |\qv''+\Gv'|^2}  d\rv  \
\Im  \chi_{\Gv \Gv'}\left[\qv'',\frac{\Pv^2}{2 M}-\frac{(\Pv-\qv)^2}{2 M}\right].
\end{split} 
\end{equation}
From this point, we consider a film of the crystal, infinite in the $x y$ plane and confined to the interval $[-L/2,L/2]$ in the $z$-coordinate.
Integrating in $\rv$ over the film, we have
\begin{equation} 
\begin{split}
\Delta E &= - \frac{8 Z^2}{\pi } \int\limits_{\qv''\in \Omega_B} d\qv'' \sum\limits_{\Gv,\Gv'}  
\int  \left[\frac{\Pv^2}{2 M}-\frac{(P_x+q''_x+G_x)^2}{2 M}-\frac{(P_y+q''_y+G_y)^2}{2 M} -\frac{(P_z-q_z)^2}{2 M}\right] \\
&\times \Theta\left[\frac{\Pv^2}{2 M}-\frac{(P_x+q''_x+G_x)^2}{2 M}-\frac{(P_y+q''_y+G_y)^2}{2 M} -\frac{(P_z-q_z)^2}{2 M}\right] \\
&\times \delta \left[  \frac{\Pv^2}{2 M}- \frac{(P_x+G_x-G'_x)^2}{2 M}- \frac{(P_y+G_y-G'_y)^2}{2 M}-\frac{(P_z-q_z-q''_z-G'_z)^2}{2 M}
   \right] \frac{ \sin[ (q_z+q''_z+G_z) L/2]}{q_z+q''_z+G_z}
    \\ 
&\times 
  \tilde{\Phi} (\Pv,t\to-\infty)  \tilde{\Phi}^* (P_x+G_x-G'_x,P_y+G_y-G'_y,P_z-q_z-q''_z-G'_z,t\to-\infty)   \\
&\times  \frac{d q_z d\Pv  }{[(q''_x+G_x)^2+(q''_y+G_y)^2+q_z^2 ]|\qv''+\Gv'|^2}  
 \Im  \chi_{\Gv \Gv'}\left[\qv'',\frac{\Pv^2}{2 M}-\frac{(P_x+q''_x+G_x)^2}{2 M}-\frac{(P_y+q''_y+G_y)^2}{2 M} -\frac{(P_z-q_z)^2}{2 M}\right].
\end{split} 
\label{L10}
\end{equation}

Let the projectile wave-packet be Gaussian    
\begin{equation}  
\tilde{\Phi} (\Pv) = \left( \frac{2}{\pi} \right)^{3 / 4}
   \sigma^{3 / 2} e^{- \sigma^2 (\Pv-\Pv_0)^2}, 
   \label{sdel}
\end{equation}
where $\sigma$ is large. In the limit $\sigma\to\infty$ Eq.~\eqref{sdel} ensures that $|\tilde{\Phi} (\Pv)|^2 \to \delta(\Pv-\Pv_0)$.
Then, due to $\tilde{\Phi}^*$ in Eq.~\eqref{L10}, only $G'_x=G_x$ and $G'_y=G_y$ contribute, which allows to rewrite Eq.~\eqref{L10} as
\begin{equation} 
\begin{split}
\Delta E &= - \frac{8 Z^2}{\pi } \int\limits_{\qv''\in \Omega_B} d\qv'' \sum\limits_{\Gv,\Gv'}  
\int   \left[\frac{\Pv^2}{2 M}-\frac{(P_x+q''_x+G_x)^2}{2 M}-\frac{(P_y+q''_y+G_y)^2}{2 M} -\frac{(P_z-q_z)^2}{2 M}\right] \\
&\times \Theta\left[\frac{\Pv^2}{2 M}-\frac{(P_x+q''_x+G_x)^2}{2 M}-\frac{(P_y+q''_y+G_y)^2}{2 M} -\frac{(P_z-q_z)^2}{2 M}\right] \\
&\times \delta \left[  \frac{P_z^2}{2 M}-\frac{(P_z-q_z-q''_z-G'_z)^2}{2 M}
   \right] \frac{ \sin[ (q_z+q''_z+G_z) L/2]}{q_z+q''_z+G_z}
    \\ 
&\times 
  \tilde{\Phi} (\Pv,t\to-\infty)  \tilde{\Phi}^* (P_x,P_y,P_z-q_z-q''_z-G'_z,t\to-\infty)   \\
&\times  \frac{d q_z d z}{[(q''_x+G_x)^2+(q''_y+G_y)^2+q_z^2 ][(q''_x+G_x)^2+(q''_y+G_y)^2+(q''_z+G'_z)^2 ]} d\Pv   \\
&\times \Im  \chi_{\Gv \Gv'}\left[\qv'',\frac{\Pv^2}{2 M}-\frac{(P_x+q''_x+G_x)^2}{2 M}-\frac{(P_y+q''_y+G_y)^2}{2 M} -\frac{(P_z-q_z)^2}{2 M}\right].
\end{split} 
\label{L11}
\end{equation}
Integrating in Eq.~\eqref{L11} over $q_z$, we  have due to the $\delta$-function

\begin{equation} 
\begin{split}
\Delta E &= - \frac{8 M Z^2}{\pi } \int\limits_{\qv''\in \Omega_B} d\qv'' \sum\limits_{\Gv,\Gv'}  
\int  \frac{1}{P_z} \left[\frac{\Pv^2}{2 M}-\frac{(P_x+q''_x+G_x)^2}{2 M}-\frac{(P_y+q''_y+G_y)^2}{2 M} -\frac{(P_z+q''_z+G'_z)^2}{2 M}\right] \\
&\times \Theta\left[\frac{\Pv^2}{2 M}-\frac{(P_x+q''_x+G_x)^2}{2 M}-\frac{(P_y+q''_y+G_y)^2}{2 M} -\frac{(P_z+q''_z+G'_z)^2}{2 M}\right]\\
&\times  \frac{ \sin[ (G_z-G'_z) L/2]}{G_z-G'_z}
  |\tilde{\Phi} (\Pv,t\to-\infty)|^2   \frac{d\Pv }{[(q''_x+G_x)^2+(q''_y+G_y)^2+q_z^2 ][(q''_x+G_x)^2+(q''_y+G_y)^2+(q''_z+G'_z)^2 ]}   \\
&\times \Im  \chi_{\Gv \Gv'}\left[\qv'',\frac{\Pv^2}{2 M}-\frac{(P_x+q''_x+G_x)^2}{2 M}-\frac{(P_y+q''_y+G_y)^2}{2 M} -\frac{(P_z+q''_z+G'_z)^2}{2 M}\right] .
\end{split} 
\label{L12}
\end{equation}

The leading term in Eq.~\eqref{L12}, proportional to $L$, is at $G'_z=G_z$, all other terms vanishing after the division by $L$. Then, after the integration over $\Pv$, we have
\begin{equation} 
\begin{split}
\lim_{L\to\infty} \frac{\Delta E}{L}  &= - \frac{4 M Z^2}{\pi P_{0 z} } \int\limits_{\qv''\in \Omega_B} d\qv'' \sum\limits_{\Gv}  
\int  \left[\frac{\Pv_0^2}{2 M}-\frac{(\Pv_0+\qv''+\Gv)^2}{2 M}\right] \\
&\times \Theta\left[\frac{\Pv_0^2}{2 M}-\frac{(\Pv_0+\qv''+\Gv)^2}{2 M}\right] 
     \frac{1}{|\qv''+\Gv|^4}     \Im  \chi_{\Gv \Gv}\left[\qv'',\frac{\Pv_0^2}{2 M}-\frac{(\Pv_0+\qv''+\Gv)^2}{2 M}\right] .
\end{split} 
\label{L13}
\end{equation}

Finally, taking account of the relation between the density-response matrix and the dielectric matrix
\begin{equation}
\epsilon^{-1}_{\Gv \Gv'}(\qv,\omega)= \delta_{\Gv \Gv'}+ \frac{4\pi}{|\qv+\Gv| |\qv+\Gv'|} \chi_{\Gv \Gv'}(\qv,\omega),
\end{equation}
we have
\begin{equation} 
\begin{split}
\lim_{L\to\infty} \frac{\Delta E}{L}  &= - \frac{ M Z^2}{ \pi^2 P_{0 z} } \int\limits_{\qv''\in \Omega_B} d\qv'' \sum\limits_{\Gv}  
\int  \left[\frac{\Pv_0^2}{2 M}-\frac{(\Pv_0+\qv''+\Gv)^2}{2 M}\right] \\
&\times \Theta\left[\frac{\Pv_0^2}{2 M}-\frac{(\Pv_0+\qv''+\Gv)^2}{2 M}\right] 
     \frac{1}{|\qv''+\Gv|^2}     \Im  \epsilon^{-1}_{\Gv \Gv}\left[\qv'',\frac{\Pv_0^2}{2 M}-\frac{(\Pv_0+\qv''+\Gv)^2}{2 M}\right] .
\end{split} 
\label{L14}
\end{equation}

\twocolumngrid

Using the symmetry property of the dielectric matrix $\epsilon^{-1}_{-\Gv',-\Gv}(-\qv,\omega)=\epsilon^{-1}_{\Gv,\Gv'}(\qv,\omega)$, we see that Eq.~\eqref{L14} proves Eq.~\eqref{QU}. In the specific case of the HEG dielectric function $\epsilon(q,\omega)$, by integrating separately over the radial and angular parts of $\qv$,  Eq.~\eqref{QU} is reduced to Eq.~\eqref{S2}.

\

\section{Details of calculations for crystals}

\label{DC}
We were carrying out calculations for real-solids with the all-electron full-potential linearised augmented-plane wave code ELK \cite{elk}.
For the realistic band-structures of semiconductors, the Tran-Blaha xc potential TB09 \cite{Tran-09} was used, and, for the uniformity, it was also used in the case of metals.

To evaluate $\chi_{s,\Gv \Gv'}(\qv,\omega)$, we were first calculating $\epsilon^{-1}_{s,\Gv \Gv'}(\qv,\omega)$ (task $180$ of the ELK code), inverting the latter matrix, then taking use of the relation
\begin{equation*}
\epsilon_{s,\Gv \Gv'}(\qv,\omega)=\delta_{\Gv \Gv'}- \frac{4\pi}{|\qv+\Gv| |\qv+\Gv'|} \chi_{s,\Gv \Gv'}(\qv,\omega).
\end{equation*}
Finally, Eqs.~\eqref{QU}--\eqref{fxc_HEG} were applied.

To account for the core-states' excitation, the promotion of the core to valence states was done.
Ideally, all electrons should be considered valence, which, however, was not always feasible due to the computational limitations. The division of electrons into the two categories used is shown in Table \ref{tab}.
The large number of empty bands is needed at larger velocities of the projectiles as well.
In general, all the parameters of the calculations were chosen such as to aim at the convergence at larger velocities, which, 
however, was not always possible. 
At lower velocities of the projectile results are well converged with respect to the parameters of the calculations.

All possible symmetries have been used to reduce the calculation time to acceptable values.

\onecolumngrid

\

\begin{table}[h!] % The 'h!' placement specifier tells LaTeX to put it "here"
\centering
\begin{tabular}{|c|c|c|c|c|c|} % | adds vertical lines; c=center, l=left, r=right
\hline % Horizontal line
Crystal & $k$- and $q$-points & Empty bands & Core electrons & Valence electrons & Dielectric matrix \\ 
\hline
Al & $16\times 16 \times 16$ & 512 & $2$ & 11 & $941\times 941$ \\
Si & $12\times 12\times 12$ & 256 & $2$ & 12 & $1471\times 1471$\\
Diamond & $16\times 16\times 16$ & 356 & 0 & 6 & $645\times 645$\\
Be & $16\times 16\times 16$ & 512  & 0 & 4 & $761\times\ 761$\\
Li & $16\times 16\times 16$ & 612 & 0 & 3 & $887\times 887$\\
Rb & $8\times 8\times 8$ & 512 &$12$ & 25 & $2731\times 2731$\\
\hline
\end{tabular}
\caption{\label{tab} Parameters of the ground-state and the dielectric matrix calculations used in producing results in Figs.~\ref{Al_te_elk}--\ref{Si_electron}.}
\label{table}
\end{table}

\pagebreak

\twocolumngrid

\section{High velocity} \label{HV}

Here we prove that any causal $f_{xc}$ does not affect the high-velocity asymptotics of the stopping power of HEG. We write down the difference between SP with inclusion of some $f_{xc}$ and the RPA one. According to Eq.~\eqref{S2}, \eqref{KS1}, and \eqref{KS2}
\begin{equation*} 
\begin{split}
&\frac{d E}{d x} -\left(\frac{d E}{d x}\right)_{RPA}= - \frac{8 Z^2}{ v^2 }  \times \\ 
& \! \! \! \! \! \int\limits_0^{2 M v} \! \!  \frac{d q}{q^3} \! \! \! \! \! \! \int\limits_0^{v q-\frac{q^2}{2 M}} 
\! \! \! \! \! \! \! \omega   \Im \! \! \left\{ \! \frac{\chi_s(q,\omega)}{1\! - \! \chi_s(q,\omega)[\frac{4\pi}{q^2} \! + \! f_{xc}(q,\omega)]} \!  - \!
\frac{\chi_s(q,\omega)}{1 \! - \! \chi_s(q,\omega) \frac{4\pi}{q^2}} \! \right\} \! d \omega,
\end{split}
\end{equation*}
or
\begin{equation*} 
\begin{split}
&\frac{d E}{d x} -\left(\frac{d E}{d x}\right)_{RPA}=  \frac{8 Z^2}{ v^2 }  \times \\ 
&  \! \! \! \! \! \int\limits_0^{2 M v} \! \!  \frac{d q}{q^3} \! \! \! \int\limits_0^{v q-\frac{q^2}{2 M}} 
\! \! \! \! \! \! \omega  \Im  \! \!  \frac{\chi_s^2(q,\omega) f_{xc}(q,\omega)}{\{1-\chi_s(q,\omega)[\frac{4\pi}{q^2}+f_{xc}(q,\omega)]\}[1-\chi_s(q,\omega) \frac{4\pi}{q^2}]} d \omega,
\end{split}
\end{equation*}
Taking  $v\to\infty$  in the upper limit of the inner integral and considering that the integrand is an even function of $\omega$, we can write asymptotically
\begin{equation} 
\begin{split}
&\frac{d E}{d x} -\left(\frac{d E}{d x}\right)_{RPA} \underset{v\to \infty}{\longrightarrow}  \frac{4 Z^2}{ v^2 }  \times \\ 
&  \! \! \! \! \! \Im  \! \! \! \! \! \int\limits_0^{2 M v} \! \! \!  \frac{d q}{q^3} \! \! \! \int\limits_{-\infty}^\infty
\! \! \!    \frac{\omega \chi_s^2(q,\omega) f_{xc}(q,\omega)}{\{1-\chi_s(q,\omega)[\frac{4\pi}{q^2}+f_{xc}(q,\omega)]\}[1-\chi_s(q,\omega) \frac{4\pi}{q^2}]} d \omega.
\end{split}
\label{CC}
\end{equation}
Furthermore, since, due to causality, the integrand in \eqref{CC} is an analytical function of $\omega$ in the upper complex half-plane, since
$\chi_s(q,\omega) \underset{|\omega|\to\infty}{\sim
} \frac{1}{|\omega|^2}$, and since $|f_{xc}(q,\omega)|$ is bounded in this half-plane \cite{Giuliani&Vignale}, by Cauchy integral theorem we conclude that the inner integral in \eqref{CC} evaluates to zero.
This proves that SP with inclusion of a $f_{xc}$ has the same high-velocity asymptotic behavior as the RPA one.

\end{document}